\documentclass[11pt]{article}
\usepackage{amsmath}
\usepackage{amssymb}
\usepackage{graphics}
\usepackage{epsf,graphicx,rotating,color}
\usepackage{bm,bbm}
\usepackage{amsfonts}
\usepackage[latin1]{inputenc}
\usepackage{graphicx}
\usepackage{graphics}
\usepackage{amsmath}
\usepackage[centerlast,small]{caption2}
\usepackage{color}

\begin{document}
\title{ Modular magnetic field on the z-direction  on a chain of nuclear spin system and
quantum Not and Controlled-Not gates.}
%\title{ Effect of the time depending magnetic field  on the z-direction during gates operation in a chain of nuclear spin quantum computer of few qubits.}
\author{G.V. L\'opez$^2$\thanks{gulopez@udgserv.cencar.udg.mx}\  and M. Avila$^1$\thanks{mavilaa@uaemex.mx}\\
\\\\$^1${\it \small Centro Universitario UAEM Valle de Chalco, UAEMex,}\\
{\it \small  Mar\'{\i}a Isabel, CP 56615, Valle de Chalco, Estado de
M\'exico, M\'exico.}  \\$^2${\it \small Departamento de F\'{\i}sica,
Universidad de Guadalajara, }\\
{\it\small  Blvd. Marcelino Garc\'{\i}a Barrag\'an 1421, esq. Calzada Ol\'{\i}mpica,}
\\{\it\small 44420
Guadalajara, Jalisco, M\'exico.}}
\date{February, 2010}
\maketitle
\begin{abstract}
 We study the simulation of a single qubit rotation and
Controlled-Not gate in a solid state one-dimensional chain of  nuclear spins
system interacting weakly through an Ising type of interaction with a modular component
of the magnetic field in the z-direction, characterized by $B_z(z,t)=Bo(z)\cos\delta t$.
These qubits are subjected to electromagnetic pulses which determine the transition in the
one or two qubits system. We use the fidelity parameter to determine the performance of the
Not (N) gate  and Controlled-Not (CNOT) gate as a function of the frequency parameter
$\delta$. We found that for $|\delta|\le 10^{-3}~MHz$, these gates still have good fidelity.
\end{abstract}
\newpage
\section{Introduction}
Almost any quantum system with at least two quantum levels may be used, in
principle, for quantum computation. This one uses qubits (quantum
bits) instead of bits to process information. A qubit is the
superposition of any two levels of the system, called $|0\rangle$
and $|1\rangle$ states, $ \Psi = C_0|0\rangle + C_1|1\rangle $ with
$|C_0|^2 + |C_1|^2 = 1$. The tensorial product of L-qubits makes up
a register of length $L$, say $|x\rangle = |i_{L-1}, . . . ,
i_0\rangle$, with $i_j = 0,\,\,1$, and a quantum computer with
L-qubits works in a $2^L$ dimensional Hilbert space, where an
element of this space is of the form     $\Psi= \sum C_x|x\rangle$,
with $\sum |C_x|^2 = 1$. Any operation with registers is done through a unitary transformation
which defines a quantum gate,  and
one of the must important result about quantum gates and quantum logical operation
is that any quantum computation can be done in terms of a single qubit unitary operation and  a Controlled-Not (CNOT) gate
 % or a single qubit unitary operation and a Controlled-Controlled-Not (CCNOT) gate since CNOT and CCNOT are universal gates \cite{basis1},\cite{basis2} \cite{basis1},\cite{basis2}. \\ \\
\noindent
Although quantum computers of few qubits \cite{boshi}-\cite{you} have been in
% operations for some time and they have been used  successfully so far,
to make serious computer calculations one may
requires a quantum computer with at least of 100-qubits registers,
and hopefully this will be achieved in a future not so far away. One
solid state quantum computer model that has been explored for
physical realization and which allows to make analytical and numerical studies of
quantum gates and protocols \cite{ZZ}  is
the one made of one-dimensional chain of nuclear spins systems

\cite{lloyd}-\cite{berman} inside  a strong magnetic field in the z-direction (with very
strong gradient in that direction)  and an RF-field in the transverse direction.
Such a model physically is unlikely to be constructed, however this represents
a good approximation for simulation of quantum algorithms and gates whose respective
results could be applied in more realistic quantum computers. Furthermore,
the approach relies in the universal character of Quantum Mechanics.
In this model, the Ising interaction is considered among first and second neighbor spins which allows to implement ideally this type of computer
up to 1000-qubits or more \cite{berman2},\cite{lopezX}. Among other gates and algorithms \cite{lopezY}, one qubit rotation and CNOT gates were study
with this quantum computer model \cite{berman3}. One of the important statement of this model
is that one keep constant the magnetic field in the z-direction at the location of each qubit. However,
this statement may be not so realistic in practice for this model or other solid state quantum computer based on spin system with very strong  axial magnetic field,  
% The main reason is that the strong magnetic field must be done with superconducting magnets which, in turns, are made of superconducting cables, and on the wires of these  cables eddy currents are induced  which can last for some time \cite{BaCra} and can can produce modulation on the magnetic field, 
and then we wonder:  if there is a magnetic field modulation
where this field change slowly with time, how these basic elements, one qubit rotation and CNOT gates, would be % affected ?.
Of course, in this case, the usual analytical approximation without field modulation is not valid anymore, and a full numerical calculation
is required to see the possible effect of this modulation on  1-qubit rotations and CNOT gates.
% In our study we will assume that the system is completely insulated from the environment, such that the important decoherence effects \cite{Ilya, Men,Zu} which normally would appear in this quantum system  is not considered.
\\\\
In this paper, we want to study this modulation effect of the magnetic field on the Not (particular case of 1-qubit rotation, or unitary operation) and CNOT quantum gates. To do this, we will assume an additional cosine time dependence on the normal z-direction of the magnetic field  and will determine, using the fidelity % \cite{chuang1,alternative,Cab} 
parameter, the minimum variation in the frequency of this modulation to keep these quantum gates elements still well defined. \\\\
\section{ Quantum Not-gate}
Consider a single paramagnetic particle with spin one-half in a magnetic field given by
\begin{equation}\label{1}
{\bf B}=(B_a\cos(\omega t),-B_a\sin(\omega t),B_0(z)\cos{\delta t})
\end{equation}
where the first two components represent the RF-field, and the third component represents the
strong magnetic field in this direction. The interaction between this particle and the magnetic field
is given by the Hamiltonian $H=-{\vec\mu}\cdot {\bf B}$, where $\vec\mu$ is the magnetic moment of
the particle which is related with the nuclear spin ${\bf\widehat S}=\hbar{\bf\hat I}$ as $\vec\mu=\gamma\hbar{\bf\hat I}$, with $\gamma$ the gyromagnetic ratio of the particle. So, the Hamiltonian is
\begin{equation}\label{2}
\widehat H=-\vec\mu\cdot {\bf B}=-\hbar\omega_o\cos{\delta t}~\hat I_z-\frac{\hbar\Omega}{2}
(\hat I_+e^{i\omega t}+\hat I_{-}e^{-i\omega t})\ ,
\end{equation}
where $\omega_o=\gamma B_0(z_o)$ ($z_o$ is the location of the particle) is the Larmor frequency, $\Omega=\gamma B_a$ is the Rabi frequency, and $\hat I_{\pm}$ represents
the ascent (descent) operator, $\hat I_{\pm}=\hat I_x\pm i\hat I_y$. If $|0\rangle$ and $|1\rangle$ are the two states of the spin one-half, one has that
\begin{equation}\label{3}
\hat I_z|i\rangle=\frac{(-1)^{i}}{2}|i\rangle\ ,\quad \hat I_+|0\rangle=|1\rangle\ ,\quad \hat I_{-}|1\rangle=|0\rangle\ .\end{equation}
% The ground state of the system is represented by $|0\rangle$, which represents the spin of the particle in the  direction of the third component of the magnetic field.  
To solve the Schr\"odinger equation,
\begin{equation}
i\hbar\frac{\partial|\Psi\rangle}{\partial t}=\widehat H|\Psi\rangle\ ,
\end{equation}
one proposes a solution of the form
\begin{equation}
|\Psi\rangle=c_o(t)|0\rangle+c_1(t)|1\rangle
\end{equation}
such that $|c_o|^2+|c_1|^2=1$ at any time. Doing this, one gets the following ordinary differential equations
\begin{subequations}
\begin{equation}
i\dot c_o=-\frac{\omega_o\cos{\delta t}}{2}c_o-\frac{\Omega}{2}c_1e^{i\omega t}
\end{equation}
and
\begin{equation}
i\dot c_1=+\frac{\omega_o\cos{\delta t}}{2}c_1-\frac{\Omega}{2}c_oe^{-i\omega t}\ .
\end{equation}
\end{subequations}
Choosing $c_0(t)=e^{i\omega t/2}d_0(t)$ and $c_1(t)=e^{-i\omega t/2}d_1(t)$ in above equations, one has
\begin{subequations}
\begin{equation}
i\dot d_0=+\frac{\omega-\omega_o\cos{\delta t}}{2}d_0-\frac{\Omega}{2}d_1
\end{equation}
and
\begin{equation}
i\dot d_1=-\frac{\omega-\omega_o\cos{\delta t}}{2}d_1-\frac{\Omega}{2}d_0
\end{equation}
\end{subequations}
which, in turns, can be written as the following uncoupled similar Mathieu  equation \cite{AA},
\begin{subequations}
\begin{equation}\label{eq:Aa}
\ddot d_0+\alpha(t)d_0=0\ \
\end{equation}
where the complex function $\alpha(t)$ is given by
\begin{equation}
\alpha(t)=\frac{1}{4}\left[\Omega^2+\omega^2\left(1-\frac{\omega_o}{\omega}\cos\delta t\right)^2\right]+i\frac{\omega_o\delta}{2}\sin\delta t\ ,
\end{equation}
\end{subequations}
and  $d_1$ is obtained from (7a),
\begin{equation}
d_1=\frac{\omega-\omega_o\cos{\delta t}}{\Omega}d_0-i\frac{2}{\Omega}\dot d_0\ .
\end{equation}
For $\delta=0$ and on resonance ($\omega=\omega_o$), one has that $\alpha=\Omega^2/4$ , and  the system oscillates between the states $|0\rangle$ and $|1\rangle$ with and angular frequency corresponding to the Rabi frequency $\Omega$, as one expected \cite{berman3}. For $\delta\not=0$ the solution of this equation is far to be trivial, and instead of solving the Eq. ($8a$), we will find  directly the numerical solution of the
system (7) with the given initial conditions.
By taking $\omega=\omega_o$ (resonant case), one expects to obtain the transition
$|0\rangle\longleftrightarrow |1\rangle$  and to get the quantum Not-gate with a phase. \\ \\
To study the performance of the quantum Not-gate as a function of the modulation frequency $\delta$ , we will calculate the fidelity parameter
at the end of a $\pi$-pulse and make the comparison  of the ideal  wave function, $\Psi_{expected}$, with the wave function resulting from our simulation, $\Psi_{sim}$.
% This fidelity parameter is the module of the complex function 
\begin{equation}\label{7}
F=\langle\Psi_{sim}|\Psi_{expected}\rangle\ ,
\end{equation}
where $|\Psi_{sim}\rangle $ is the state obtained from numerical simulations, and $|\Psi_{expected}\rangle$ is the ideal expected state. for the initial condition $|\Psi_o\rangle=|0\rangle$, of course, the fidelity coincide with the coefficient $|c_1|^2$.
At this point we want to stress that we define $|F|^2$ in this way due that any quantum gate or algorithm
is represented by the final wave function of the quantum system. Ideally,
if the quantum gate is fully realizable this wave function is represented by
$|\Psi_{expected}\rangle$. However, the non resonant transitions and
the error systems (modulation) make that the resulting wave function of the
complete simulation is given by $|\Psi_{sim}\rangle$. In this way, the fidelity
is a measure of the good operation of gates and algorithms. On the other
hand, there is another measurement for the calculation of the the distance between two states and this 
is the so called
Uhlmann-Josza fidelity \cite{uhlmann}. However, in Ref. \cite{alternative} it has been shown
that Eq. (10) is a lower bound for the Uhlmann-Josza fidelity. Such a result 
favors the present results.

Fig. 1a and Fig. 1b  show the behavior of the fidelity and the probabilities as a function of the parameter $\delta$ at the end of a $\pi$-pulse, $\tau=\pi/\Omega$. We have used the parameters (units $2\pi~MHz$) $\Omega=0.1$ and $\omega_o=200$. The RF-frequency has been chosen equal to the resonant frequency $\omega=\omega_o$.  As one can see, for $\delta\le 0.2\times 10^{-3}MHz$ we can have a very well defined quantum Not-gate. \\\\
%
%\begin{figure}[H]\label{Fig1}
\begin{figure}[H]
 \centering
\includegraphics[width=8.5in,height=6in]{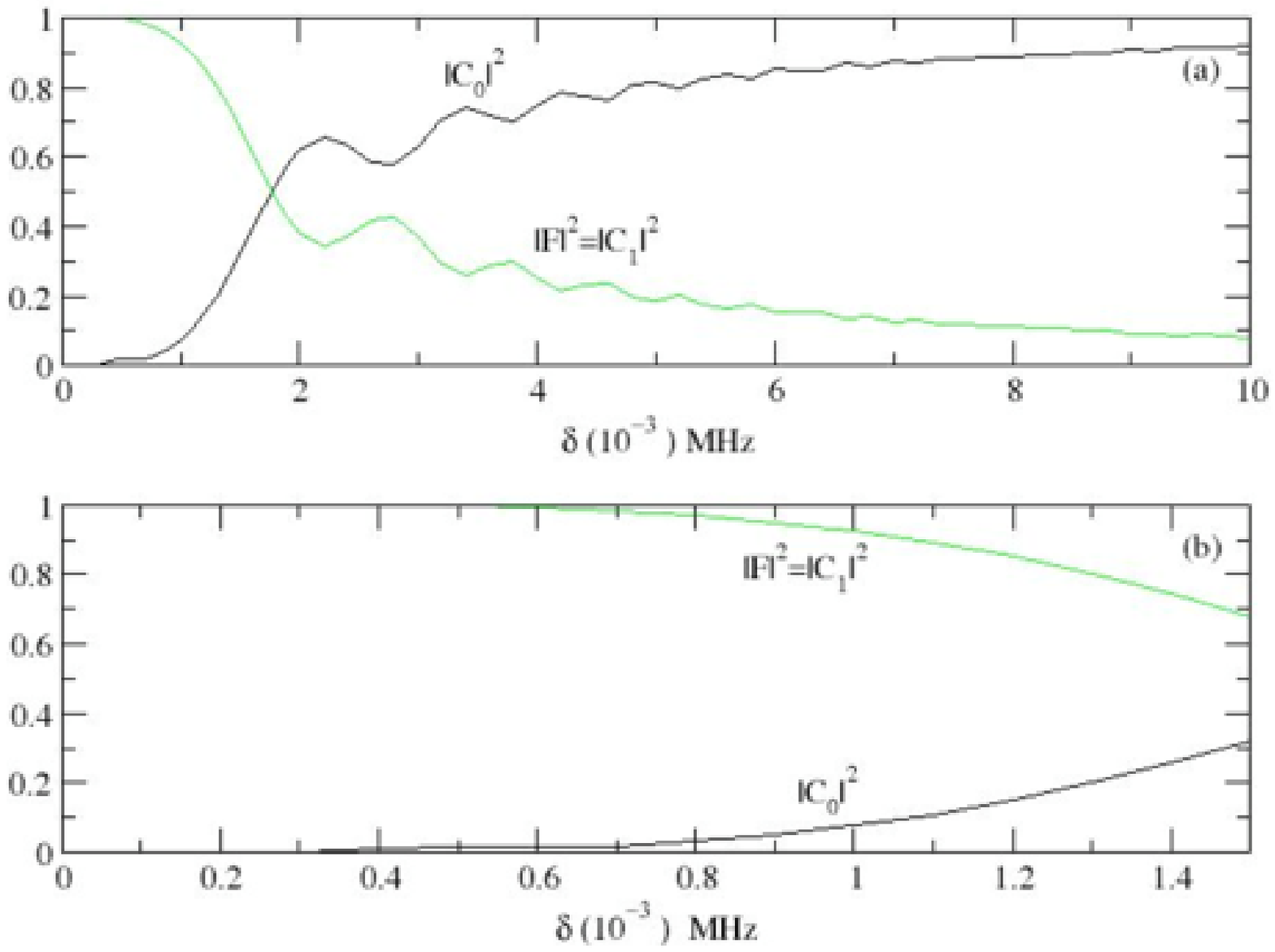}
 \caption{ Quantum Not-gate: (a) Global behavior (b) Local behavior with respect to $\delta$.}
\label{Fig1}
\end{figure}
\section{Two qubits  model and quantum CNOT gate}
Fig.2 shows  two paramagnetic nuclear particles of spin one-half (qubits)
subjected to a magnetic field of Eq. (1),  making and angle $\cos\theta=\sqrt{3}/2$ to eliminate the dipole-dipole interaction between them. The interaction of the magnetic field with the
qubits is carried out through the coupling with their dipole magnetic moment
$\vec \mu_i=\gamma {\bf S} _i ~~(i=1,\,\,2$), where
$\gamma$ is the gyromagnetic ratio and ${\bf \widehat S}_i$ is the spin of the ith-nucleon  (${\bf \widehat S}=\hbar {\bf\hat I}$). The interaction energy is given by
\begin{eqnarray}
\widehat H&=&-\vec \mu_1 \cdot {\bf  B}_1-\vec \mu_2 \cdot {\bf B}_2+\hbar J
\hat I_z^{(1)} \hat I_z^{(2)} \nonumber  \\
&=& \widehat H_0-\frac{\hbar \Omega}{2}\left(\hat I_+^{(1)}e^{i\omega
t}+\hat I_-^{(1)}e^{-i\omega t}+\hat I_+^{(2)}e^{i\omega
t}+\hat I_-^{(2)}e^{-i\omega t}\right),
\end{eqnarray}
\noindent
where $J$ is the coupling constant of interaction between nearest neighboring spins, $\Omega=\gamma B_a$  is  the Rabi frequency, $\widehat H_0$ is the part of Hamiltonian which is diagonal in the basis
$\{|i_1i_o\rangle\}_{i_j=0,1}$ and is given by
\begin{equation}
\widehat H_0= -\hbar \left(\omega_1 \hat I_z^{(1)}+\omega_2
\hat I_z^{(2)}\right) \cos{\delta t}+\hbar J\hat I_z^{(1)}\hat I_z^{(2)}.
\end{equation}
where $\omega_i$ are the Larmor's frequencies which are defined as
\begin{equation}
\omega_i=\gamma B_0(z_i) \quad i=1,2
\end{equation}
with $z_i$ being the z-location of the ith-qubit. The eigenvalues of $\widehat H_0$ on the above basis for $\delta=0$ are
\begin{equation}
\begin{split}
E_{00}&=-\frac{1}{2}\{\omega_1+\omega_2-\frac{1}{2}J\}\quad
E_{01}=-\frac{1}{2}\{\omega_1-\omega_2+\frac{1}{2}J\}\\
E_{10}&=-\frac{1}{2}\{-\omega_1+\omega_2+\frac{1}{2}J\}\quad
E_{11}=-\frac{1}{2}\{-\omega_1-\omega_2-\frac{1}{2}J\}
\end{split}
\end{equation}
\begin{figure}[H]
 \centering
 \includegraphics[width=8.5cm,height=6cm]{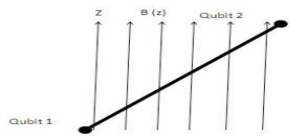}
 \caption{ Two qubits configuration.}
\label{Fig2}
\end{figure}
\noindent
% The ground state of the system is denoted by $|00\rangle$ which corresponds to the case of having both spins parallel in the direction of the third component of the magnetic field.
By doing  $\omega=(E_{11}-E_{10})/\hbar=\omega_2-J/2$, one gets the resonant transition which
defines the CNOT operation $|10\rangle\longleftrightarrow |11\rangle$ with a phase involved ($e^{i\pi/2}$), where the left qubits is the control and the right one is the target. To solve the Schr\"odinger equation,
\begin{equation}
i\hbar \frac{\partial| \Psi\rangle}{\partial t}=\widehat H|\Psi\rangle\ ,
\end{equation}
we can  assume that the wave function can be written as
\begin{equation}
\Psi=C_{00}(t)|00\rangle+C_{01}(t)|01\rangle
+C_{10}(t)|10\rangle+C_{11}(t)|11\rangle
\end{equation}
such that  $\sum |C_{ij}|^2=1$.
Thus, we arrive to the following system of complex-couple ordinary differential equations
\begin{subequations}
\begin{equation}
i \dot{C}_{00}=-\frac{1}{2}\left((\omega_1+\omega_2)\cos{\delta t}-\frac{1}{2}J\right)C_{00}-\frac{\Omega}{2}\left(C_{01}+C_{10}\right)e^{i\omega t}
\end{equation}
\begin{equation}
i\dot{C}_{01}=-\frac{1}{2}\left((\omega_1-\omega_2)\cos{\delta t}+\frac{1}{2}J\right)C_{01}-\frac{\Omega}{2}\left(C_{00}e^{-i\omega t}+C_{11}e^{i\omega t}\right)
\end{equation}
\begin{equation}
i \dot{C}_{10}=-\frac{1}{2}\left((\omega_2-\omega_1)\cos{\delta t}+\frac{1}{2}J\right)C_{10}-\frac{\Omega}{2}\left(C_{00}e^{-i\omega t}+C_{11}e^{i\omega t}\right)
\end{equation}
\begin{equation}
i \dot{C}_{11}=-\frac{1}{2}\left(-(\omega_1+\omega_2)\cos{\delta t}-\frac{1}{2}J\right)C_{11}-\frac{\Omega}{2}\left(C_{01}+C_{10}\right)e^{-i\omega t}.
\end{equation}
\end{subequations}
Doing the transformation $C_{00}=e^{i\omega t/2}D_{00}$, $C_{01}=e^{-i\omega t/2}D_{01}$, $C_{10}=e^{-i\omega t/2}D_{10}$, and $C_{11}=e^{-i3\omega t/2}D_{11}$, one gets rid of the fast oscillations and gets the following equations for the coefficients $D's$:
\begin{subequations}
\begin{equation}
i \dot{D}_{00}=-\frac{1}{2}\left((\omega_1+\omega_2)\cos{\delta t}-\frac{1}{2}J-\omega\right)D_{00}-\frac{\Omega}{2}\left(D_{01}+D_{10}\right)
\end{equation}
\begin{equation}
i\dot{D}_{01}=-\frac{1}{2}\left((\omega_1-\omega_2)\cos{\delta t}+\frac{1}{2}J+\omega\right)D_{01}-\frac{\Omega}{2}\left(D_{00}+D_{11}\right)
\end{equation}
\begin{equation}
i \dot{D}_{10}=-\frac{1}{2}\left((\omega_2-\omega_1)\cos{\delta t}+\frac{1}{2}J+\omega\right)D_{10}-\frac{\Omega}{2}\left(D_{00}+D_{11}\right)
\end{equation}
\begin{equation}
i \dot{D}_{11}=-\frac{1}{2}\left(-(\omega_1+\omega_2)\cos{\delta t}-\frac{1}{2}J+3\omega\right)D_{11}-\frac{\Omega}{2}\left(D_{01}+D_{10}\right).
\end{equation}
\end{subequations}
We solve numerically these equation, and for $\delta=0$ and $\omega=\omega_2-J/2$, a full transition will occur between the states $|10\rangle$ and $|11\rangle$. Note that one has $C_{ij}(0)=D_{ij}(0)$ and $|C_{ij}(t)|^2=|D_{ij}(t)|^2$. For $\delta\not=0$,  we consider two initially conditions cases: Digital case, where the initial condition is given by
\begin{subequations}
\begin{equation}
|\Psi_o\rangle=|10\rangle\ ,
\end{equation}
that is $C_{00}(0)=0, \quad C_{01}(0)=0,\quad
C_{10}(0)=1,\quad
C_{11}(0)=0$.
Superposition case, where the initial condition is
\begin{equation}
|\Psi_o\rangle=\sqrt{\frac{2}{10}}|00\rangle+\frac{1}{\sqrt{10}}|01\rangle+\sqrt{\frac{6}{10}}|10\rangle+
\frac{1}{\sqrt{10}}|11\rangle\ .
\end{equation}
\end{subequations}
% These two initial states can be gotten from our ground state $|00\rangle$ by applying  it Hadamard and/or CNOT gates which is not the point in our study. So, we are assuming that these initial states are given, and we use a simple state and a superposition state to cover  a general situation. 
For our simulation, we use the following parameters (units $2\pi~MHz$)    $\Omega=0.1$, $\omega_1=100$, $\omega_2=110$, and $J=10$.
The RF-frequency chosen is the resonant frequency  $\omega=\omega_2-J/2$, and applying  a $\pi$-pulse, $\tau=\pi/\Omega$, we should get the respective CNOT transition $|10\rangle \longleftrightarrow |11\rangle$.  Fig. 3 shows the behavior of the probabilities and the fidelity as a function of the parameter $\delta$ at the end of the $\pi$-pulse and for the digital case.  Fig. 4 shows the same as before but for the superposition case. This case is more stable (the fidelity decays more slowly than the digital case) due to non zero contribution to the terms $C_{00}$ and $C_{01}$ which always contribute with the same constant probability $3/10$.
\newpage
\section{Conclusion}
For a quantum computer model of a chain of qubits in a magnetic field where its z-component
 varies with respect the time, we have studied the Not and Controlled-Not gate behavior as a
 function of the frequency $\delta$ of variation of this component.  In general, one can say that for $\delta\le 10^{-3}MHz$ these quantum gates remain well defined with a fidelity very close to one. This small value in $\delta$ means that it is enough to consider a first order in taylor expansion of the cosine function in Eq. (1).
 We have seen that the fidelity for the superposition case is more stable than the digital case due to the contribution to the fidelity parameter of the other no zero states involved in the dynamics. Of course, this safety region, defined by $\delta$, for these quantum gates does not mean safety for a full quantum algorithm, which is under  studied.
\begin{figure}[H]
 \centering
  \includegraphics[width=10.5cm,height=8cm]{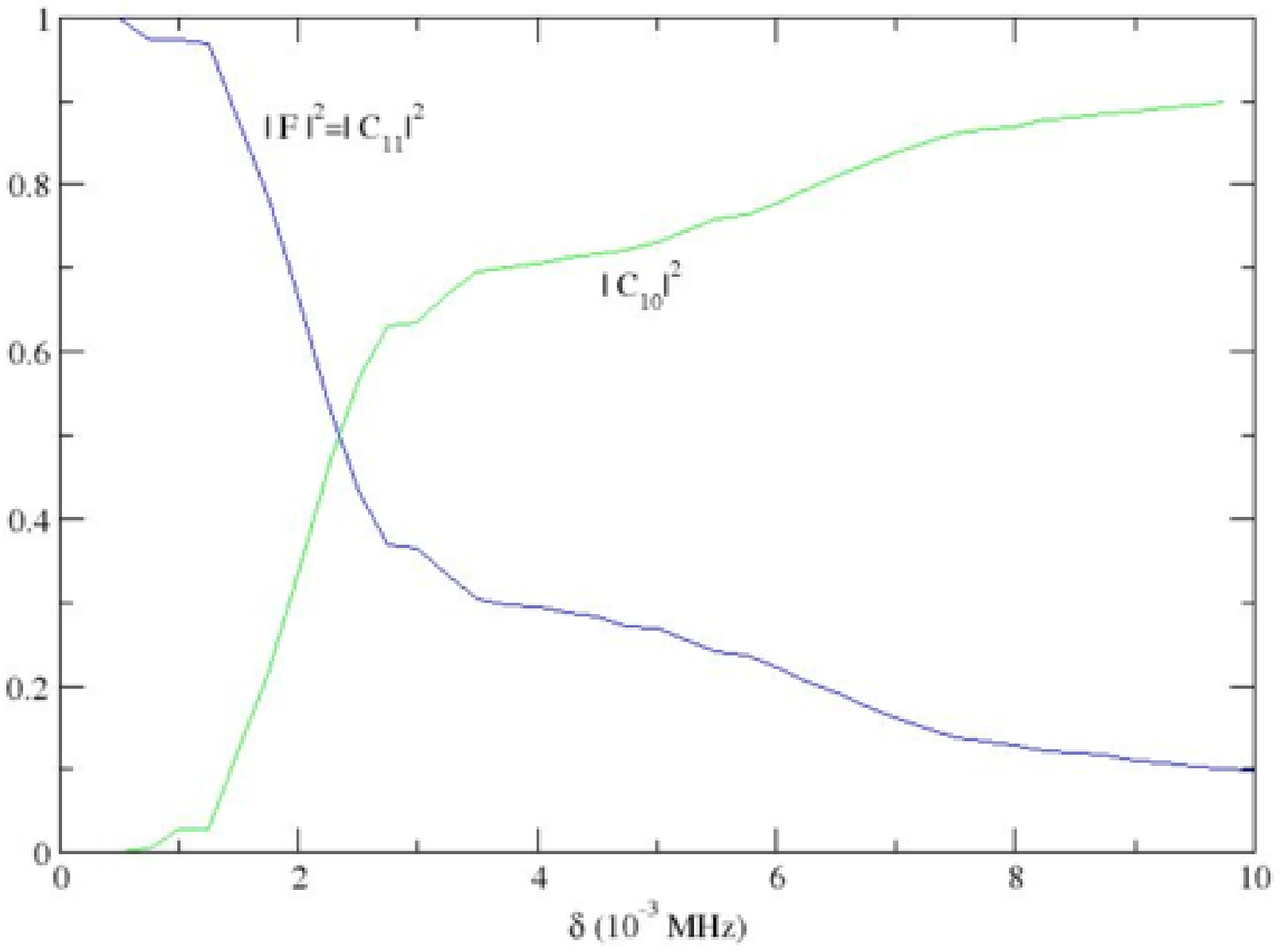}
 \caption{ CNOT behavior, digital case.}
\label{Fig3}
\end{figure}
\begin{figure}[H]
 \centering
  \includegraphics[width=6.5cm,height=4cm]{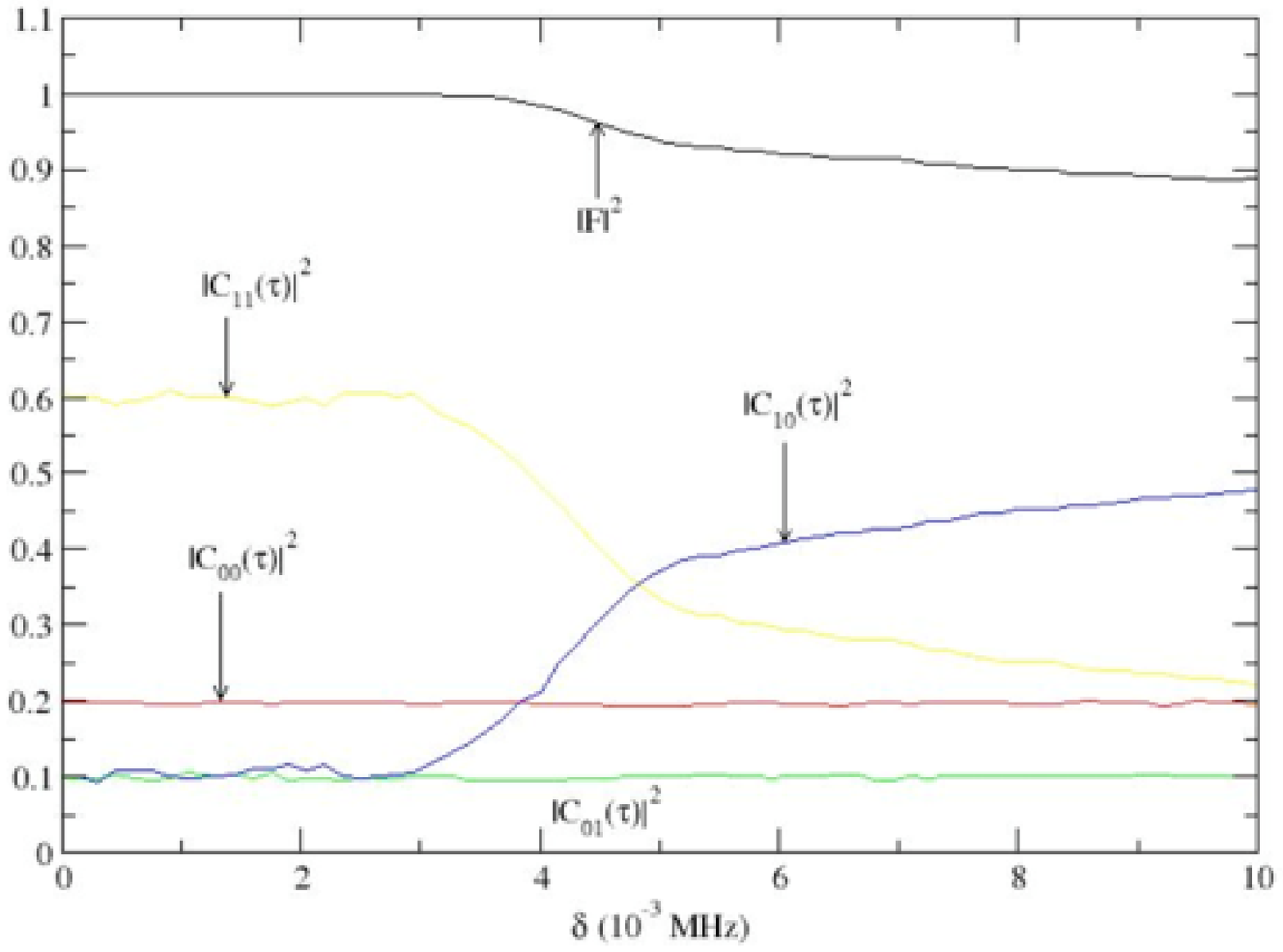}
 \caption{ CNOT behavior, superposition case.}
\label{Fig4}
\end{figure}
\section { Acknowledgments}
We want to thank UAEMex for the grant  2594/2008U.

\end{document}